\documentclass[a4paper,11pt]{article}

\usepackage{amsfonts,amscd,amssymb,amsmath,amsthm,latexsym}
\usepackage{verbatim}
\def\bea{\begin{eqnarray}}
\def\eea{\end{eqnarray}}
\def\ba{\begin{array}}
\def\ea{\end{array}}

\newcommand{\FF}{\mathcal{F}}

\def\dst{\displaystyle}

\numberwithin{equation}{section}

\date{}

\begin{document}

%=================================================================
% Full title of the paper (Capitalized)
\title{New  Cosmological Solutions of a Nonlocal  Gravity Model}

% Authors, for the paper (add full first names)
\author{\small {Ivan Dimitrijevic$^{1}$, Branko Dragovich$^{2,3}$, Zoran Rakic$^{1}$ and Jelena Stankovic$^{4}$} \\
\small {$^{1}$University of Belgrade, Faculty of Mathematics,  Studentski trg 16,  11158 Belgrade, Serbia} \\
\small {$^{2}$University of Belgrade, Institute of Physics,  11080  Belgrade, Serbia; dragovich@ipb.ac.rs} \\
\small {$^{3}$Mathematical Institute of the Serbian Academy of Sciences and Arts, 11000 Belgrade, Serbia} \\
\small {$^{4}$University of Belgrade, Teacher Education Faculty, Kraljice Natalije 43, 11000  Belgrade, Serbia}}

\maketitle

\abstract{A nonlocal gravity model \eqref{eq2.1}
was introduced and considered recently, \cite{dimitrijevic10},
and two exact  cosmological solutions
in flat space were presented. The first solution is related to some radiation effects generated by nonlocal dynamics on  dark energy background, while
the second one is a nonsingular time symmetric bounce. In the present paper we investigate other possible exact cosmological solutions and find some the new ones
in nonflat space. Used nonlocal gravity dynamics can change background topology. To solve the corresponding  eqations of motion, we first look for a
solution of the eigenvalue problem
$\Box (R -4\Lambda) = q\ (R - 4\Lambda) .$  We also discuss possible extension of this model with nonlocal operator symmetric under
$\Box \longleftrightarrow \Box^{-1}$ and its connection  with another interesting
nonlocal gravity model. }

%\begin{document}
%%%%%%%%%%%%%%%%%%%%%%%%%%%%%%%%%%%%%%%%%%

\section{Introduction}

The current  Standard Model of Cosmology (SMC) \cite{robson},  also known as $\Lambda$CDM model,  assumes General Relativity  (GR) \cite{wald} as
theory of the gravitational interaction at all cosmic space-time scales -- galactic and cosmological.
According to this  model, at current cosmic time the universe approximately contains 68\% of dark energy (DE),  27\% of dark matter (DM) and
5\% of visible  matter.  By $\Lambda$CDM model, dark matter is responsible for
observational dynamics inside and between galaxies, while dark energy  causes accelerated expansion of the universe.
$\Lambda$CDM model also asserts that DE corresponds to the cosmological constant and that
DM is in a cold state. In the last few decades  many efforts were done to confirm existence of DM and
DE in the sky or in the laboratory experiments, but they are not discovered, and their existence still remains
hypothetical. A brief review of recent investigations of  DM and DE is presented in \cite{oks}.

 Due to its significant phenomenological achievements and  beautiful theoretical properties, GR is considered as one of the basic modern physical theories \cite{ellis}.
 For example, GR  describes dynamics of the Solar system very well.  Many important phenomena  were also predicted and observationally confirmed:
 deflection of light near the Sun,  black holes,  as well as gravitational light redshift, lensing, and  waves.
However, GR as a theory of gravitation has not been
verified at the galactic and cosmological scales. Despite remarkable successes, GR solutions  for the black holes
and  the beginning of the universe contain  singularities. In addition, from quantization point of view, GR is a nonrenormalizable theory.
Note also that every other physical theory has its domain of validity which is usually constrained by space-time scale,
complexity of the system under consideration, or by some parameters. There is no {\it a priori} reason that GR is an exception and
should be theory of gravitation from the Planck scale to the universe as a whole.   Taking  into account all these remarks it follows that
general relativity is not a final gravitational theory  and that investigation of its extension is needed, e.g., see
\cite{faraoni,clifton,nojiri,nojiri1,novello,antoniadis,nissimov,djordjevic} and references therein.

 Since it is not invented so far  a new  physical  principle that could say in which direction extend GR,
 there are many approaches to its modification (for a review, see  \cite{faraoni,clifton,nojiri,nojiri1,novello}).
 One of the current and attractive approaches to general relativity modification is its nonlocal extension, see e.g.
 \cite{deser,woodard,maggiore,biswas1,biswas2,biswas3,biswas5,biswas4,biswas6,kolar} and
 \cite{dragovich0,koshelev1a,koshelev1b,koshelev1,koshelev2,koshelev3,koshelev4,eliz,koivisto}.
The idea behind nonlocality is that dynamics of the gravitational field may depend not only
on its first and second space-time derivative but also on all higher derivatives. It means that the Einstein-Hilbert action should be extended
by an additional nonlocal term that
contains the d'Alembert-Beltrami operator $\Box$ which is mainly employed in two ways: (i) using an analytic expansion
$F (\Box) = \sum_{n=0}^{+\infty} f_n \Box^n$, or (ii)  including in some manner operator $\Box^{-1}$ \cite{deser,woodard,maggiore,jovanovic},
and its higher powers.

  The modification of type (i) comes from ordinary and $p$-adic string theory, see \cite{dragovich1} and references therein. This type of nonlocality improves quantum
renormalizability \cite{stelle,modesto1,modesto2}. Nonlocal gravity models  of type (i) that have attracted much attention
  are given by  action
\begin{equation} \label{eq1.1}
S = \frac{1}{16 \pi G}\int_{\mathcal{M}} \sqrt{-g}\, \big(R - 2\Lambda + P(R)\, \mathcal{F}(\Box)\, Q(R)\big)\, d^4x ,
\end{equation}
where $\mathcal{M}$ is a four-dimensional pseudo-Riemannian manifold of signature $(-, +, +, +)$ with metric $(g_{\mu\nu})$,
$P(R)$ and $Q(R)$ are some differentiable functions of  scalar curvature $R$, $\Lambda$ is the cosmological constant, and $\mathcal{F} (\Box) = \sum_{n=0}^{+\infty} f_n \Box^n$.
To  better see effects of nonlocal modification of GR in its geometrical sector, action \eqref{eq1.1} intentionally does not contain matter term.
Derivation of equations of motion that are related to nonlocal gravity  \eqref{eq1.1} is a difficult task, and for details we refer to our paper \cite{dimitrijevic9}, see also \cite{biswas4}.

  Action \eqref{eq1.1} is rather general and contains several simple nonlocal extensions of GR.
   $P(R)=Q(R)=R$ is a case that has attracted the most attention, see
  \cite{biswas1,biswas2,koshelev1a,koshelev1b} and
  \cite{dimitrijevic1,dimitrijevic2,dimitrijevic3,dimitrijevic4,dimitrijevic5,dimitrijevic6,dimitrijevic7,dimitrijevic8}.
 It incudes  also nonlocal extension of the Starobinsky $R^2$ inflation model \cite{koshelev1,koshelev2}.
 This kind of nonlocal investigation started in \cite{biswas1,biswas2} and is an attempt to find nonsingular bouncing
solution of the  singularity problem in standard cosmology. It is worth mentioning  an interesting model
when $P(R)=Q(R)= \sqrt{R - 2\Lambda}$, which contains
cosmological solution $a(t) = A t^{\frac{2}{3}} e^{\frac{\Lambda}{14}t^2}$ that mimics  an interference between dark matter ($t^{\frac{2}{3}}$)
and dark energy
($e^{\frac{\Lambda}{14}t^2}, \ \Lambda > 0$) in flat space ($k = 0$).  Explored cosmological parameters are in  good agreement with $\Lambda$CDM
data, see \cite{PLB}.

This paper is devoted to the further  investigation of  the nonlocal gravity model which is given by $P(R) = Q(R) = R - 4 \Lambda $,
 and presented in \cite{dimitrijevic10}.
The nonlocal term $(R - 4\Lambda)\ \mathcal {F} (\Box) \ (R - 4\Lambda)$ appears as a generalization of $R\ \mathcal {F} (\Box) \ R .$ This model is also of interest  as the limit case of model $P(R) = Q(R) = \sqrt{R - 2\Lambda}$
for $ |R| \ll |2\Lambda|$, see Section 2. In the paper \cite{dimitrijevic10} we investigated the exact cosmological
solutions for $ \Lambda \neq 0, \, k=0$: $a_1(t) = A \sqrt{t} e^{\frac{\Lambda}{4}t^2} ,$ and $a_2(t) = A e^{\Lambda\,t^2}.$  The first
solution mimics an interplay between  dark energy and radiation. The second solution is a nonsingular bounce one and an even function of cosmic time.
 In this paper we consider new cosmological solutions with scale factors of the
 two forms:  $a(t)= \Big(\alpha e^{\lambda t} +\beta e^{-\lambda t}\Big)^\gamma$ and
 $a(t)= \big(\alpha \cos \lambda t + \beta \sin \lambda t\big)^\gamma ,$ where $\gamma$ is an arbitrary real parameter.

The paper is organized as follows. In Section 2, the concrete nonlocal gravity model is set up and some general properties of the relevant
equations of motion are presented.  Section 3 contains consideration of various aspects of the corresponding cosmological solutions: brief review of two previous results, relevant eigenvalue problem and detailed analysis related to finding of the new exact
cosmological solutions.  Discussion and conclusions are presented in Section 4.

\section{Gravity Model with Additional Nonlocal Term \\ $(R-4 \Lambda)\ \mathcal{F}(\Box)\ (R -4 \Lambda)$}

Nonlocal gravity model under consideration is given by  action

\begin{equation} \label{eq2.1}
S = \frac{1}{16 \pi G} \int d^4 x \ \sqrt{-g}\ \big(R- 2 \Lambda +  (R-4 \Lambda)\ \mathcal{F}(\Box)\ (R -4 \Lambda) \big) ,
\end{equation}
 where   $\Box = \nabla_{\mu}\nabla^{\mu}= \frac{1}{\sqrt{-g}}\, \partial_\mu \, (\sqrt{-g}\, g^{\mu\nu}\, \partial_\nu )$ is the
 d'Alembert-Beltrami operator  on the corresponding gravity background and  $\mathcal{F}(\Box) = \sum_{n=1}^{+\infty} f_n \ \Box^n $ is
 nonlocal operator with all higher order space-time derivatives. Formally, \eqref{eq2.1} gets from  \eqref{eq1.1} taking $P(R) = Q(R) = R - 4  \Lambda$
 and $f_0 = 0 .$  However, \eqref{eq2.1} can be also derived from action
\begin{equation} \label{eq2.2}
S = \frac{1}{16 \pi G} \int d^4 x \ \sqrt{-g}\ \big(R- 2 \Lambda +  \sqrt{R-2 \Lambda}\ \mathcal{F}(\Box)\ \sqrt{R-2 \Lambda} \big)
\end{equation}
which also belongs to the class of nonlocal models \eqref{eq1.1}. In fact, let us start from action \eqref{eq2.2} and consider expansion of
$\sqrt{R-2 \Lambda} = \sqrt{-2\Lambda}\ \sqrt{1 -\frac{R}{2\Lambda}}$ in powers of $\frac{R}{2\Lambda},$ where $|R| \ll |2\Lambda|$. Then let
us  take approximation linear in $\frac{R}{2\Lambda}$, i.e. one obtains $\sqrt{R-2 \Lambda} \simeq \sqrt{-2\Lambda}\ (1 -\frac{R}{4\Lambda})$.
By this way, nonlocal term in \eqref{eq2.2} becomes
\begin{equation} \label{eq2.3}
  \sqrt{R-2 \Lambda}\ \mathcal{F}(\Box)\ \sqrt{R-2 \Lambda}  \simeq -\frac{1}{8\Lambda} (R-4 \Lambda)\ \mathcal{F}(\Box)\ (R -4 \Lambda),
\end{equation}
where factor $-\frac{1}{8\Lambda}$ can be included in nonlocal operator $\mathcal{F}(\Box)$ by its redefinition.
At the same time, the first term $R - 2\Lambda = \sqrt{R-2 \Lambda}\ \sqrt{R-2 \Lambda}$ remains unchanged in the linear approximation.

As it is already mentioned in Introduction, nonlocal gravity model \eqref{eq2.2} is very interesting and promising. It is natural nonlocal generalization
of the de Sitter model
\begin{align} \label{eq2.5}
  S_0 =  \frac{1}{16 \pi G} \int d^4 x \ \sqrt{-g}\ (R - 2 \Lambda),
\end{align}
where generalization obtains in the following way:
\begin{equation}\label{eq2.6}
  R-2\Lambda = \sqrt{R-2 \Lambda}\ \sqrt{R-2 \Lambda} \rightarrow  \sqrt{R-2 \Lambda}\ F(\Box) \sqrt{R-2 \Lambda} .
\end{equation}
Nonlocal operator $F(\Box)$ in \eqref{eq2.6} is  $F(\Box) = 1 + \mathcal{F}(\Box) = 1 + \sum_{n=1}^{+\infty} f_n\ \Box^n$. Nonlocal de Sitter model (also called nonlocal square
root gravity \cite{PLB}) \eqref{eq2.2}  contains two exact scale factors:
\begin{equation}\label{eq2.7}
  a(t) = A t^{\frac{2}{3}} e^{\frac{\Lambda}{14}t^2} , \quad  a(t) = A e^{\frac{\Lambda}{6}\,t^2} , \quad k=0.
\end{equation}
The first solution in \eqref{eq2.7} mimics  an interference between dark matter expansion ($t^{\frac{2}{3}}$) and dark energy acceleration
($e^{\frac{\Lambda}{14}t^2}, \ \Lambda > 0$) in flat space ($k = 0$), and calculated cosmological quantities are in  good agreement with standard model of cosmology, see details in \cite{PLB}. The second solution in \eqref{eq2.7} is an example of nonsingular bounce at cosmic time $t = 0$.

\subsection{Equations of Motion}

The next step in investigation of nonlocal gravity model \eqref{eq2.1} is finding the corresponding equations of motion (EOM). It is done for a class of
models \eqref{eq1.1}, that contains \eqref{eq2.1}, and derivation is presented in \cite{dimitrijevic9}.

According to \cite{dimitrijevic9},  the EOM for nonlocal gravity model \eqref{eq1.1} have the following form:
 \begin{align} \label{eq2.8}
 \hat G_{\mu\nu} = G_{\mu\nu} +\Lambda g_{\mu\nu} - \frac 12 g_{\mu\nu} P(R) \mathcal{F}(\Box) Q(R) + R_{\mu\nu}W -K_{\mu\nu}W + \frac 12 \Omega_{\mu\nu} = 0,
 \end{align}
 where
 \begin{align}
 &W =  P'(R)\ \mathcal{F}(\Box)\ Q(R) + Q'(R)\ \mathcal{F}(\Box) P(R) , \quad K_{\mu\nu} = \nabla_\mu \nabla_\nu - g_{\mu\nu}\Box , \label{eq2.9a} \\
  &\Omega_{\mu\nu} =  \sum_{n=1}^{+\infty} f_n \sum_{\ell=0}^{n-1} S_{\mu\nu}(\Box^\ell P(R), \Box^{n-1-\ell} Q(R)) ,  \label{eq2.9b} \\
  &S_{\mu\nu} (A, B) = g_{\mu\nu} \big(\nabla^\alpha A \ \nabla_\alpha B + A \Box B \big) - 2\nabla_\mu A\ \nabla_\nu B , \label{eq2.9c}
\end{align}
and $P', \ Q'$ denote derivative of $P, \ Q$ with respect to $R$.

 It is clear that EOM  \eqref{eq2.8} are very complicated comparing them to their local (Einstein) counterpart $G_{\mu\nu} +\Lambda g_{\mu\nu} = 0$.
Finding any solutions of  \eqref{eq2.8} is not an easy task. However, in the sequel of this article we will see how one can find some exact cosmological solutions   when $P(R) =Q(R) = R- 4 \Lambda ,$ i.e. in the nonlocal gravity model \eqref{eq2.1}.

 First, let us consider the case when $Q(R) = P(R)$. Then EOM (\ref{eq2.8}) reduce to:
\begin{align} \label{eq2.10a}
  &G_{\mu\nu}+ \Lambda g_{\mu\nu} - \frac{g_{\mu\nu}}{2} P(R) \FF (\Box) P(R) + R_{\mu\nu} W - K_{\mu\nu} W + \frac 12 \Omega_{\mu\nu} = 0 , \\
&W = 2 P'(R)\ \FF (\Box)\ P(R), \quad K_{\mu\nu} = \nabla_\mu \nabla_\nu - g_{\mu\nu}\Box , \label{eq2.10b} \\
&\Omega_{\mu\nu} =  \sum_{n=1}^{+\infty} f_n \sum_{\ell=0}^{n-1} S_{\mu\nu}(\Box^\ell P, \Box^{n-1-\ell} P) \label{eq2.10c} \\
&S_{\mu\nu}(\Box^\ell P, \Box^{n-1-\ell} P) =\sum_{\ell=0}^{n-1} \Big( g_{\mu\nu} \big(\nabla^\alpha \Box^\ell P(R) \ \nabla_\alpha \Box^{n-1-\ell} P(R)
+ \Box^\ell P(R)  \Box^{n-\ell} P(R) \big) \nonumber \\
&\qquad \qquad  - 2\nabla_\mu \Box^\ell P(R)\ \nabla_\nu \Box^{n-1-\ell} P(R) \Big) . \label{eq2.10d}
\end{align}

 The further significant  simplification of EOM can be obtained if $P(R)$ is an eigenfunction of the corresponding d'Alembert-Beltrami operator $\Box$,  i.e. if holds
\begin{align} \label{eq2.11}
\Box P(R) = q\,  P (R), \ \quad  \ \FF (\Box) P(R) = \FF (q)\, P(R) ,
\end{align}
where $q = \zeta \Lambda$  ($\zeta$ dimensionless parameter) is an eigenvalue.   Note that parameter $q$ must have the same dimensionality as $\Box$, where dimension of $\Box$ is $T^{-2}$ in natural units ($\bar{h} = c = 1$). Hence, $q$ has to be proportional to $\Lambda$, since there is only the cosmological constant $\Lambda$ in  the above EoM  with dimension as $\Box.$ Moreover, $q = \zeta \Lambda$ naturally appears in all concrete cases and there is no need for   a new constant in this nonlocal gravity model without matter.
 Then
\begin{align}
  &W = 2 \FF(q)P' P,  \quad \FF(q) = \sum_{n=1}^{+\infty} f_n \ q^n , \label{eq2.12a} \\
  &\Omega_{\mu\nu} = \FF'(q) S_{\mu\nu}(P,P), \label{eq2.12b} \\
  &G_{\mu\nu}+ \Lambda g_{\mu\nu} + \FF (q)\Big(  2(  R_{\mu\nu}
  -   K_{\mu\nu} ) P P'   - \frac{g_{\mu\nu}}{2}  P^2 \Big) \nonumber \\
  &+ \frac 12 \FF'(q) S_{\mu\nu}(P,P) = 0. \label{eq2.12c}
\end{align}
Both expressions in \eqref{eq2.12a} are evident. Equality \eqref{eq2.12b} obtains as follows:
\begin{align}
\Omega_{\mu\nu} &= \sum_{n=1}^\infty f_n \sum_{\ell =0}^{n-1} S_{\mu\nu}\big(\Box^\ell P , \Box^{n-1-\ell} P \big)
      =\sum_{n=1}^\infty f_n \sum_{\ell =0}^{n-1} q^{n-1} S_{\mu\nu} \big(P , P \big) \label{eq2.13a} \\
   &=\sum_{n=1}^\infty f_n \ n  q^{n-1} S_{\mu\nu} \big(P , P \big)  = \mathcal{F}' (q) S_{\mu\nu} \big(P , P \big) . \label{eq2.13b}
\end{align}

 Finally, take $P= R-4\Lambda$. Then  $P\,P' = P= R-4\Lambda $ and EOM become
 \begin{align}
 &G_{\mu\nu}+ \Lambda g_{\mu\nu} + \FF (q)\Big(G_{\mu\nu} + R_{\mu\nu} - 2 \nabla_\mu \nabla_\nu + 2 g_{\mu\nu} (\Lambda + q)  \Big) (R - 4\Lambda) \nonumber \\
  &+ \frac 12 \FF'(q) S_{\mu\nu}(R - 4\Lambda,R - 4\Lambda) = 0. \label{eq2.14}
\end{align}

 In some cases there is solution when $\FF' (q) = 0$  and then problem \eqref{eq2.14} reduces to:
 \begin{align}
 &\FF' (q) = 0,\ \  \mbox{and}\ \ \label{eq2.15}  \\
 &G_{\mu\nu}+ \Lambda g_{\mu\nu} + \FF (q)\Big(G_{\mu\nu} + R_{\mu\nu} - 2 \nabla_\mu \nabla_\nu + 2 g_{\mu\nu} (\Lambda + q)  \Big) (R - 4\Lambda) = 0 .
    \label{eq2.16}
  \end{align}
  In finding cosmological solutions, we start from equations \eqref{eq2.14}.

%%%%%%%%%%%%%%%%%%%%%%%%%%%%%%%%%%%%%%%%%%
\section{Cosmological Solutions}

In this section we are mainly  interested in finding and investigation of some new exact cosmological solutions of nonlocal gravity model \eqref{eq2.1}.

Since the universe is homogeneous and isotropic at large cosmic scales, hence its evolution satisfies the
Friedmann-Lema\^{i}tre-Robertson-Walker (FLRW) metric
\begin{align}
ds^2 = - dt^2 + a^2(t)\left(\frac{dr^2}{1-k r^2} + r^2 d\theta^2 + r^2 \sin^2 \theta d\phi^2\right), \quad (c=1) , \, \, k= 0, \pm 1 , \label{eq3.1}
\end{align}
where $a(t)$ is the cosmic scale factor that  contains information on expansion (or contraction) and $k$ is the constant curvature parameter.

 The  d'Alembert-Beltrami operator $\Box$, the Hubble parameter $H$ and the Ricci scalar $R$ for the FLRW metric are:
\bea  \label{eq3.2} \dst{\ba{ll}
\dst{\Box = -\frac{\partial^2}{\partial t^2} - 3 H(t)\ \frac{\partial}{\partial t} }, \qquad & \dst{H (t) = \frac{\dot{a}}{a} } \\[13pt]
\dst{R(t) = 6 \Big(\frac{\ddot{a}}{a} + \big(\frac{\dot{a}}{a}\big)^2 + \frac{k}{a^2}\Big)} , \qquad & \dst{\dot{a} = \frac{\partial a}{\partial t} .} \ea}
\eea

Since the universe is homogeneous and isotropic there are only two independent equations of motion \eqref{eq2.14}. It is convenient to use
the trace and $00$-component of \eqref{eq2.14}:
\begin{align}
&(R - 4 \Lambda) \big[\mathcal{F}(\zeta\Lambda) (8 + 6 \zeta)\Lambda -1\big] + \frac{1}{2} \mathcal{F}'(\zeta\Lambda)\ S(R-4\Lambda,R-4\Lambda) = 0, \label{eq3.3} \\
&G_{00} - \Lambda + \mathcal{F}(\zeta\Lambda) \Big(2 R_{00} + \frac{1}{2} R - 2 \partial_0^2 - 2(1 +\zeta)\Lambda\Big)(R - 4\Lambda) \nonumber \\
&+ \frac{1}{2} \mathcal{F}'(\zeta\Lambda)\ S_{00}(R-4\Lambda,R-4\Lambda) = 0,
\label{eq3.4}
\end{align}
where $S(R-4\Lambda,R-4\Lambda) = g^{\mu\nu} S_{\mu\nu} (R-4\Lambda,R-4\Lambda)$ and equality $\Box (R - 4\Lambda)= q (R - 4\Lambda)
 = \zeta \Lambda (R - 4\Lambda)$ has been taken into account.
According to \eqref{eq3.4} we have to use
\begin{align}\label{eq3.5}
R_{00} = - 3\, \frac{\ddot{a}}{a} \, , \qquad   G_{00} = 3\, \frac{\dot{a}^2  + k}{a^2}.
\end{align}

%%%%%%%%%%%%%%%%%%%%%%%%%%%%%%%%%%%%%%%%%%%%%%%%%%%%

Note that EOM \eqref{eq2.14} can be rewritten in the form of general relativity
\begin{equation}
\hat{G}_{\mu\nu} = G_{\mu\nu}+ \Lambda g_{\mu\nu} - 8 \pi G \hat{T}_{\mu\nu} = 0 \, ,  \qquad \nabla^\mu \hat{G}_{\mu\nu} = 0,  \label{eq3.6}
\end{equation}
where $\hat{T}_{\mu\nu}$ is the corresponding effective energy-momentum tensor.
The related Friedmann equations to \eqref{eq3.6} are
\begin{equation}
\frac{\ddot{a}}{a} = - \frac{4\pi G}{3} (\bar{\rho} + 3 \bar{p}) + \frac{\Lambda}{3} \,,  \quad \qquad \frac{\dot{a}^2  + k}{a^2} = \frac{8\pi G}{3} \bar{\rho}
+ \frac{\Lambda}{3} \,,   \label{eq3.7}
\end{equation}
where $\bar{\rho}$ is an effective energy density and $\bar{p}$ is an effective pressure of the universe.
 The corresponding equation of state is
\begin{equation}
\bar{p} (t) = \bar{w}(t) \, \bar{\rho} (t) ,  \label{eq3.8}
\end{equation}
where $\bar{w}(t)$ is a dimensionless parameter that may depend on time.

It is worth noting that the Minkowski space ($a(t) =$ const.,\ $R = \Lambda = k = 0$) is also a solution of EOM \eqref{eq3.3} and \eqref{eq3.4}.

%%%%%%%%%%%%%%%%%%%%%%%%%%%%%%%%%%%%%%%%%%%%%%%%%%%%%%%%%%%%%%%
\subsection{Two Previous Exact Solutions}

In order to have more complete insight into $a(t)$ solutions of nonlocal  model \eqref{eq2.1}, we want first briefly review previously
found two nontrivial solutions \cite{dimitrijevic10} and after that present new exact solutions.

We found  the following exact cosmological solutions, $\Lambda \neq 0,\;k=0$:
\bea\label{eq3.1.1}&& \hspace{-5mm}
 a_1(t)  = A\ t^{\frac{1}{2}}\ e^{\frac{\Lambda}{4} t^2} , \qquad\ \ \Box \big(R - 4 \Lambda\big) =  - 3\Lambda \big(R - 4 \Lambda \big),\; \\[2pt]
 \label{eq3.1.2}&& \hspace{-5mm}  a_2(t)  =
 A\  e^{\Lambda t^2} ,\qquad\qquad  \Box (R - 4\Lambda) = - 12 \Lambda (R - 4 \Lambda).\eea
 We explicitly found expression for $R(t)$, $H(t)$, solved the corresponding  eigenvalue problems and EOM for both $a_1(t)$ and  $a_2(t)$, and also
  found  constraints on the nonlocal operator function $ \FF (\Box)$:
 \bea
   && \hspace{-15mm} (a_1):\ \ \mathcal{F} \big(- 3\Lambda \big) = - \frac{1}{10 \Lambda} \,,  \qquad   \mathcal{F}' \big(- 3\Lambda \big) = 0 \, ,
    \qquad \Lambda \neq 0 ,  \label{eq3.1.3} \\[2pt]
    && \label{eq3.1.4} \hspace{-15mm} (a_2): \ \
    \mathcal{F} \big(- 12\Lambda \big) = - \frac{1}{64 \Lambda} \,,  \qquad   \mathcal{F}' \big(- 12\Lambda \big) = 0 \, , \qquad \Lambda \neq 0,
\eea that are simply satisfied by
\bea  && \hspace{-25mm} (a_1): \ \
\mathcal{F}(\Box) = \frac{\Box}{30\Lambda^2} \exp{\left(\frac{\Box}{3\Lambda} + 1\right)} , \label{eq3.1.5}\\[2pt]
&& \hspace{-25mm} (a_2): \ \
\mathcal{F}(\Box) = \frac{\Box}{768\Lambda^2} \exp{\left(\frac{\Box}{12\Lambda} + 1\right)} , \label{3.1.6}\eea
respectively.

  The solution of the effective Friedmann equations were also found in both cases, and consequently  the equations of  state are:
\bea
&& \label{3.15b} \hspace{-15mm} (a_1): \ \  \bar{w} = \frac{\bar{p}(t)}{\bar{\rho} (t)}=\frac{1-6\Lambda t^{2} - 3\Lambda^2 t^4  }{3 + 2\Lambda t^{2} + 3\Lambda^2 t^4  } \quad\to\quad   \begin{cases}
-1 ,  \, \, \, \; \;  t \to \infty \\[5pt]
\;\dst{\frac{1}{3}} \;, \quad  t \to  0 .
\end{cases}\\[3pt]
&& \label{3.2.9} \hspace{-15mm} (a_2): \ \
\bar{w} =  \frac{ - 12\Lambda t^2 - 3}{ 12\Lambda t^2  - 1} \to  \begin{cases}
-1 ,  \, \, t \to \infty \\
3 , \quad  t \to  0 .
\end{cases}
\eea

 $(a_1)$: \ \  This  solution may be relevant to the early radiation dominant universe   and to its late accelerated expansion.
 The solution mimics interference between expansion with radiation $a(t) = A \sqrt{t}$ and a dark energy $a(t) = A e^{\frac{\Lambda}{4} t^2} , \ \
 \Lambda > 0 .$\vspace{1mm}

 $(a_2)$: \ \  The solution $a_2(t) = A e^{\Lambda t^2}$ is an even function of cosmic time and presents an example of the nonsingular bounce.

%%%%%%%%%%%%%%%%%%%%%%%%%%%%%%%%%%%%%%%%%%%%%%%%%%%%%%%%%%%%%%%%%%%%%

We are now going to explore the existence of new cosmological solutions with scale factors $a(t)$ similar to those well known in the de Sitter local model \eqref{eq2.5}, but with time dependent  scalar curvature $R (t)$ so that $\Box (R-4\Lambda) = q (R-4\Lambda),$ where $q \neq 0 .$
In fact, scale factors in the form $a(t) = (\alpha e^{\lambda t} + \beta e^{-\lambda t})^\gamma$  and $ a(t) = (\alpha \cos \lambda t + \beta \sin \lambda t)^\gamma$ will be investigated. We will see that such solutions exist and their significance will be discussed in Section 4.

\subsection{Eigenvalue Problem for  New Cosmological Solutions}

Let us consider the scale factor
\begin{equation}
  a(t) = (\alpha e^{\lambda t} + \beta e^{-\lambda t})^\gamma, \label{eq3.2.1}
\end{equation}
and an eigenvalue problem
\begin{equation}
  \Box (R-4\Lambda) = q (R-4\Lambda) = \zeta \Lambda (R-4\Lambda) , \label{eq3.2.2}
\end{equation}
for a dimensionless constant $\zeta$ that will be determined later.
The equality \eqref{eq3.2.2} can be expanded into
\begin{align}
& \left(\beta +\alpha  e^{2 \lambda  t}\right)^2 \left(A_0 + A_1 e^{2\lambda t} + A_2 e^{4 \lambda  t} \right) \nonumber \\
&+2 \left(\alpha  e^{\lambda  t}+\beta  e^{-\lambda t}\right)^{2
   \gamma } \Big(B_0 + B_1 e^{2\lambda t} + B_2e^{4 \lambda  t} +B_3e^{6 \lambda  t} + B_4 e^{8 \lambda  t} \Big) =0, \label{eq3.2.3}
\end{align}

where

\begin{equation} \label{eq3.2.4}
\begin{aligned}
A_0 &=3k\beta^2 \left(q-2 \gamma^2\lambda^2\right), \\
A_1 &=6k \alpha  \beta  \left(2 (\gamma -2) \gamma  \lambda^2 +q\right), \\
A_2 &=3k \alpha^2 \left(q-2 \gamma^2\lambda^2\right),
\end{aligned}
\end{equation}
and
\begin{equation} \label{eq3.2.5}
\begin{aligned}
B_0 &= \beta^4 q \left(3 \gamma^2 \lambda^2-\Lambda \right), \\
B_1 &= 2 \alpha  \beta^3 \left(6 \gamma  \left(6 \gamma^2 -7 \gamma +2\right)
   \lambda^4 +q \left(3 \gamma  \lambda^2 -2 \Lambda \right)\right), \\
B_2 &= -6 \alpha^2 \beta^2 \left(4 \gamma  \left(6 \gamma^2 -11 \gamma +4\right) \lambda^4+ q  \left(\gamma^2 \lambda^2 -2 \gamma  \lambda^2+\Lambda \right)\right), \\
B_3 &= 2 \alpha^3 \beta \left(6 \gamma  \left(6 \gamma^2 -7 \gamma +2\right) \lambda^4 +q \left(3 \gamma  \lambda^2 -2 \Lambda \right)\right), \\
B_4 &= \alpha^4 q \left(3 \gamma^2 \lambda^2-\Lambda \right).
\end{aligned}
\end{equation}

In the case $\alpha\beta = 0$, i.e.  $\alpha=0$ or $\beta=0$,  the eigenvalue problem  $\Box (R-4\Lambda) = q (R-4\Lambda)$ has nontrivial solution
in the following two cases:
\begin{enumerate}
\item $k=0$, $\Lambda = 3 \gamma^{2}\lambda^{2}$
\item $k\neq 0$, $q=2 \gamma^{2}\lambda^{2}$, $\Lambda = 3 \gamma^{2}\lambda^{2}$.
\end{enumerate}
When $\alpha\beta\neq 0$ then functions $e^{2\lambda t}$ and $\left(\alpha  e^{\lambda  t}+\beta  e^{-\lambda t}\right)^{2 \gamma }$ are linearly independent.
In this case we can split equation \eqref{eq3.2.3} into
%Functions $e^{2\lambda t}$ and $\left(\alpha  e^{\lambda  t}+\beta  e^{-\lambda t}\right)^{2 \gamma }$ are linearly independent
%if $\gamma \neq 1$ or $\gamma = 1$ and $\beta \neq 0$.

\begin{align}
  A_0 =A_1 = A_2 =0, \qquad B_0=B_1 = B_2= B_3 = B_4 =0.\label{eq3.2.6}
\end{align}

 The previous equations  \eqref{eq3.2.6}  are satisfied in the following two cases:
\begin{align}
&1. \qquad \gamma=1,\ q=2 \lambda^{2},\ \Lambda=3 \lambda^{2},\ k \in \{0,-1,1\} ,  \label{eq3.2.6a}  \\ %[1pt]
&2. \qquad \gamma=\frac{1}{2},\ \Lambda=\frac{3}{4} \lambda^{2},\ k=0.  \label{eq3.2.6b}
\end{align}

Hence, the only two possibilities for parameter $\gamma$ are: $\gamma =1$ and $\gamma = \frac{1}{2}$.

Now, let us consider the scale factor
\begin{equation}
  a(t) = (\alpha \cos \lambda t + \beta \sin \lambda t)^\gamma , \label{eq3.2.7}
\end{equation}
and the corresponding eigenvalue problem
\begin{equation}
  \Box (R-4\Lambda) = q (R-4\Lambda). \label{eq3.2.8}
\end{equation}
Similarly as in the previous case, if we replace $\lambda$ by $\mathrm{i} \lambda$ in the scale factor
$a(t) = (\alpha e^{\lambda t} + \beta e^{-\lambda t})^\gamma$ we obtain that the eigenvalue problem \eqref{eq3.2.8} has solution in the following two cases:
\begin{align}
&1. \qquad \gamma=1,\ q= -2 \lambda^{2},\ \Lambda= -3 \lambda^{2},\ k \in \{0,-1,1\} ,  \label{eq3.2.9a}  \\ %[1pt]
&2. \qquad \gamma=\frac{1}{2},\ \Lambda= -\frac{3}{4} \lambda^{2},\ k=0.  \label{eq3.2.9b}
\end{align}

As result of the solution of eigenvalue problem \eqref{eq3.2.2} we obtained not only concrete eigenvalue $q$ but also possible values of
$\gamma$ and $\lambda$ for the cosmic scale factor of the form \eqref{eq3.2.1} and \eqref{eq3.2.7}. In fact,
 we have found that  nonlocal gravity model \eqref{eq2.1} may have the following new cosmological solutions:
 \begin{align}
 &a_3(t) = \alpha\, e^{\sqrt{\frac13\,\Lambda}\, t}+\beta e^{-\sqrt{\frac13\,\Lambda}\, t}, \qquad \Lambda\geq 0 \, , \label{eq3.2.10a} \\
 &a_4(t) = \left(\alpha e^{2\sqrt{\frac13\,\Lambda}\, t} +\beta e^{-2\sqrt{\frac13\,\Lambda}\, t}\right)^{\frac 12} , \quad \Lambda\geq 0, \label{eq3.2.10b} \\
 &a_5(t) = \alpha\, \cos \sqrt{-\frac13\,\Lambda}\, t +\beta \sin\, \sqrt{-\frac13\,\Lambda}\, t \, ,  \qquad \Lambda\leq 0  ,  \label{eq3.2.10c} \\
 &a_6(t) = \left(\alpha \cos 2\sqrt{-\frac13\,\Lambda}\, t +\beta \sin 2\sqrt{-\frac13\,\Lambda}\, t\right)^{\frac 12} , \quad \Lambda\leq 0 . \label{eq3.2.10d}
\end{align}

By additional requirement that scale factors
\eqref{eq3.2.10a} -- \eqref{eq3.2.10d} satisfy equations of motion \eqref{eq3.3} and \eqref{eq3.4} gives possibility to determine values of $\alpha$ and $\beta$,
fix curvature constant $k$ and obtain  constraints on  $\mathcal{F} (\Box)$ and $\mathcal{F}' (\Box)$.  In the following four subsections we give more details.

%%%%%%%%%%%%%%%%%%%%%%%%%%%%%%%%%%%%%%%%%%
\subsection{Cosmological Solution of the Form $a_3(t)= \alpha e^{\sqrt{\frac{\Lambda}{3}}\;t} +\beta e^{-\sqrt{\frac{\Lambda}{3}}\;t}$}

In this case we have
\begin{align}
&\dot{a}(t) = \sqrt{\frac{\Lambda}{3}}\Big(\alpha e^{\sqrt{\frac{\Lambda}{3}}\;t} -\beta e^{-\sqrt{\frac{\Lambda}{3}}\;t}\Big),
\quad \ddot{a}(t) =\frac{\Lambda}{3} a(t) , \label{eq3.3.1a} \\
&R(t) =4 \Lambda + (6k -8 \Lambda \alpha\beta)a(t)^{-2} , \label{eq3.3.1b} \\
&H(t) =\sqrt{\frac{\Lambda}{3}}\Big(1-2 \beta e^{-\sqrt{\frac{\Lambda}{3}}\;t}a(t)^{-1}\Big) , \label{eq3.3.1c} \\
&R_{00} =-\Lambda, \ \quad G_{00} = \Lambda + (3k -4 \Lambda \alpha\beta)a(t)^{-2} .  \label{eq3.3.1d}
\end{align}
The corresponding  eigenvalue problem has the following solution:
\begin{align} \label{eq3.3.2}
\Box (R - 4\Lambda) = \frac{2}{3} \Lambda (R - 4 \Lambda) , \quad  \mathcal{F}(\Box) (R -4 \Lambda) = \mathcal{F}\big(\frac{2}{3} \Lambda\big) (R -4 \Lambda) .
\end{align}

 Using the solution of  eigenvalue problem \eqref{eq3.3.2}, the trace and $00$-component of EOM are:
\begin{align}
  &\left(4 \alpha  \beta  \frac \Lambda3-k\right) \Big(T_0+ T_1 e^{2 \sqrt{\frac\Lambda3}  t} + T_2 e^{4 \sqrt{\frac\Lambda3}  t} +
  T_3 e^{6 \sqrt{\frac\Lambda3}  t}+ T_4 e^{8 \sqrt{\frac \Lambda3}  t} \Big) =0,  \label{eq3.3.3a}\\
  &\left(4 \alpha  \beta  \frac \Lambda3-k\right) \Big(Z_0 + Z_1 e^{2 \sqrt{\frac \Lambda3}  t} + Z_2 e^{4\sqrt{\frac \Lambda3} t}+ Z_3 e^{6 \sqrt{\frac \Lambda3}  t}
  + Z_4 e^{8 \sqrt{\frac\Lambda3} t} \Big) =0, \label{eq3.3.3b}
\end{align}
where
\begin{equation} \label{eq3.3.4}
  \begin{aligned}
    T_0 &= \beta ^4 \left(12 \Lambda \FF\left(\frac 23\Lambda\right)-1\right),\\
    T_1 &= 4 \alpha  \beta ^3 \left(12\Lambda \FF\left(\frac 23\Lambda \right) -1 \right),\\
    T_2 &= -6 \alpha  \beta   \left(\alpha  \beta\left(1-12\Lambda
   \FF\left(\frac 23\Lambda\right) \right) -\frac{16}3 \Lambda \FF'\left(\frac 23\Lambda\right) \left(k-4 \alpha  \beta  \frac\Lambda3\right)\right), \\
    T_3 &= 4 \alpha ^3 \beta  \left(12\Lambda \FF\left( \frac 23 \Lambda \right) -1 \right),\\
    T_4 &= \alpha^4 \left(12\Lambda \FF\left(\frac 23 \Lambda\right)-1\right),
  \end{aligned}
\end{equation}

and

\begin{equation} \label{eq3.3.5}
  \begin{aligned}
    Z_0 &= \beta^4 \left(1-12 \Lambda \FF\left(\frac 23 \Lambda\right)\right),\\
    Z_1 &= 2 \beta ^2 \left(2 \alpha  \beta -6 \Lambda  \FF'\left(\frac 23\Lambda\right) \left(k-4 \alpha
   \beta  \frac \Lambda3\right)+3 \FF\left(\frac 23 \Lambda\right) \left(k-4 \alpha  \beta \Lambda \right)\right),\\
    Z_2 &= 6 \alpha  \beta  \left(\alpha  \beta +\frac 43 \Lambda \FF'\left(\frac 23\Lambda\right) \left(k-4 \alpha  \beta  \frac\Lambda3\right) + 2
  \FF\left(\frac 23 \Lambda\right) \left(k-2 \alpha  \beta  \Lambda\right)\right), \\
    Z_3 &= 2 \alpha ^2 \left(2 \alpha  \beta -6 \Lambda \FF'\left(\frac 23 \Lambda \right) \left(k-4 \alpha  \beta  \frac \Lambda3 \right)+3 \FF\left(\frac 23 \Lambda\right) \left(k-4 \alpha  \beta  \Lambda \right)\right),\\
    Z_4 &= \alpha^4 \left(1-12 \Lambda \FF\left(\frac 23 \Lambda\right)\right).
  \end{aligned}
\end{equation}

These two equations are polynomials in $e^{2\sqrt{\frac \Lambda3} t}$. Both equations are clearly satisfied if $\alpha\beta = \frac {3k}{4\Lambda}$.On the other hand, if $\alpha\beta \neq \frac {3k}{4\Lambda}$ it remains to solve the system of equations
\begin{align}
T_0 = T_1 = T_2 = T_3 = T_4 =0, \qquad Z_0 = Z_1 = Z_2 = Z_3 = Z_4 =0.  \label{eq3.3.6}
\end{align}

 Equations of motion are satisfied in the following three nontrivial cases:
\begin{align}
(i): \quad  &\alpha \beta = \frac {3k}{4\Lambda} ,  \label{eq3.3.5a}  \\  %[1pt]
(ii): \quad &\alpha \beta=0 , \ \mathcal{F}(\frac 23\Lambda)=\frac{1}{12 \Lambda} , \ \mathcal{F}\;'(\frac 23\Lambda) =
\frac{1}{24 \Lambda^{2}} , \ \ k\neq 0,  \label{eq3.3.5b} \\  % [1pt]
(iii): \quad &\alpha \beta = -\frac {k}{4\Lambda} , \  \mathcal{F}(\frac 23\Lambda)=\frac{1}{12 \Lambda} , \ \mathcal{F}\;'(\frac 23\Lambda) =0 .
 \label{eq3.3.5c}
\end{align}

\noindent$(i)$:\quad In the first case, we have $R(t)= 4 \Lambda$. For $k=0$ we have $\alpha\beta=0$ and consequently, $a(t)\sim e^{\pm\sqrt{\frac\Lambda3}\;t}$.
Also since $\Lambda>0,$ $a(t) =\sqrt{\frac{3}{\Lambda}} \cosh \sqrt{\frac\Lambda3}\;t$ requires $k=+1$, while $a(t) =\sqrt{\frac{3}{\Lambda}}
\sinh \sqrt{\frac\Lambda3}\;t$ if $k=-1$.

\noindent$(ii)$:\quad In the second case $\alpha=0$ or $\beta=0$. For $\alpha=0$ we have $a(t)=\beta e^{-\sqrt{\frac{\Lambda}{3}}\;t}$ and
$R(t)=6k a(t)^{-2} + 4 \Lambda$. Analogously, for $\beta=0$ we have $a(t)=\alpha e^{\sqrt{\frac{\Lambda}{3}}\;t}$ and $R(t)=6k a(t)^{-2} + 4 \Lambda$.

\noindent$(iii)$:\quad In the third case, we have $R(t)=4 \Lambda + 8k a(t)^{-2}$. If  $k=-1$ there is $\varphi$ such that
\begin{align*}
\alpha+\beta= \frac{1}{\sqrt{\Lambda}}\cosh \varphi,\qquad\qquad
\alpha-\beta= \frac{1}{\sqrt{\Lambda}}\sinh \varphi.
\end{align*}
Now, we can transform scale factor $a(t)= \alpha e^{\sqrt{\frac{\Lambda}{3}}\;t} +\beta e^{-\sqrt{\frac{\Lambda}{3}}\;t}$ to
\begin{align}
a(t)=\frac{1}{\sqrt{\Lambda}}\cosh (\varphi+ \sqrt{\frac{\Lambda}{3}} t) , \quad k= -1 .  \label{eq3.3.6}
\end{align}
If  $k= +1$ there is  such  $\varphi$ that
\begin{align*}
\alpha+\beta= \frac{1}{\sqrt{\Lambda}}\sinh \varphi,\qquad\qquad \alpha-\beta= \frac{1}{\sqrt{\Lambda}}\cosh \varphi.
\end{align*}
Consequently, we can transform scale factor $a(t)= \alpha e^{\sqrt{\frac{\Lambda}{3}}\;t} +\beta e^{-\sqrt{\frac{\Lambda}{3}}\;t}$ to
\begin{align}
a(t)=\frac{1}{\sqrt{\Lambda}}\sinh (\varphi+ \sqrt{\frac{\Lambda}{3}} t).  \label{eq3.3.7}
\end{align}

 Effective energy density and pressure are given by:
\begin{align}
  \bar{\rho} =\frac{3}{8 \pi G}(k-\frac{4}{3} \Lambda\alpha\beta)a(t)^{-2}, \qquad\qquad
  \bar{p} = -\frac{1}{8 \pi G}(k-\frac{4}{3} \Lambda\alpha\beta)a(t)^{-2}.   \label{eq3.3.8}
\end{align}
For $k \neq \frac{4}{3}\Lambda \alpha\beta$ the corresponding $\bar{w}$ parameter is $\bar w =-\frac{1}{3}$.

\subsection{Cosmological Solutions of the Form $ a_4(t) = \left(\alpha e^{2\sqrt{\frac{\Lambda}{3}}\, t} +
\beta e^{-2\sqrt{\frac{\Lambda}{3}}\, t}\right)^{\frac 12}$}

According to  solution \eqref{eq3.2.6b} of the related eigenvalue problem, in this case $k = 0$. The corresponding
Ricci scalar is
\begin{equation}
  R=4\Lambda.  \label{eq3.4.1}
\end{equation}
The EOM yield the  condition
\begin{equation}
  \alpha \beta = 0. \label{eq3.4.2}
\end{equation}
Hence, there are only  solutions  $a(t)\sim e^{\pm\sqrt{\frac{\Lambda}3}\;t}$, what is just we have in the Einstein theory of gravity.
Since the corresponding eigenvalue is zero, i.e. $\Box (R - 4\Lambda) = 0$, solutions of the form $ a_4(t) = \left(\alpha e^{2\sqrt{\frac{\Lambda}{3}}\, t} +
\beta e^{-2\sqrt{\frac{\Lambda}{3}}\, t}\right)^{\frac 12}$ are trivial at the classical level from the point of view of nonlocal
gravity model under consideration.

\subsection{Cosmological Solutions of the Form $a_5(t)= \alpha \cos \sqrt{-\frac{\Lambda}{3}}\;t + \beta \sin \sqrt{-\frac{\Lambda}{3}}\;t$}

In this case we have
\begin{align} \label{eq3.5.1}
&\dot{a}(t) = \sqrt{-\frac{\Lambda}{3}}( \beta \cos \sqrt{-\frac{\Lambda}{3}}\;t -\alpha \sin \sqrt{-\frac{\Lambda}{3}}\;t), \quad \ddot{a}(t) =\frac{\Lambda}{3} a(t) ,  \\
&R(t) =4 \Lambda + 6 (k-(\alpha^{2}+ \beta^{2})\frac{\Lambda}{3}a(t)^{-2}, \\
&H(t) =\sqrt{-\frac{\Lambda}{3}}( \beta \cos \sqrt{-\frac{\Lambda}{3}}\;t -\alpha \sin \sqrt{-\frac{\Lambda}{3}}\;t)a(t)^{-1}, \label{3.2.1a} \\
&R_{00} =-\Lambda \ \quad G_{00} = 3(k - \frac{\Lambda}{3}( \beta \cos \sqrt{-\frac{\Lambda}{3}}\;t -\alpha \sin \sqrt{-\frac{\Lambda}{3}}\;t)^{2})        a(t)^{-2}.
\end{align}
The corresponding  eigenvalue problem has the same solution as in the previous case \eqref{eq3.3.2}, i.e.
\begin{align} \label{eq3.5.2}
\Box (R - 4\Lambda) = \frac{2}{3} \Lambda (R - 4 \Lambda) , \quad  \mathcal{F}(\Box) (R -4 \Lambda) = \mathcal{F}(\frac{2}{3} \Lambda) (R -4 \Lambda) .
\end{align}

Trace and $00$-component of equations of motion read
\begin{align}
  &\left(k-\frac \Lambda3(\alpha ^2+\beta ^2)\right) \Big(U_0+ U_1 e^{2 \mathrm{i} \sqrt{-\frac\Lambda3}  t} + U_2 e^{4\mathrm{i}
  \sqrt{-\frac\Lambda3}  t} + U_3 e^{6 \mathrm{i} \sqrt{-\frac\Lambda3}  t}+ U_4 e^{8\mathrm{i} \sqrt{-\frac \Lambda3}  t} \Big) =0, \label{3.5.3a}\\
  &\left(k-\frac \Lambda3(\alpha ^2+\beta ^2)\right) \Big(V_0 + V_1 e^{2\mathrm{i} \sqrt{-\frac \Lambda3}  t} +
  V_2 e^{4\mathrm{i}\sqrt{-\frac \Lambda3} t}+ V_3 e^{6 \mathrm{i} \sqrt{-\frac \Lambda3}  t}
  + V_4 e^{8\mathrm{i} \sqrt{-\frac\Lambda3} t} \Big) =0, \label{3.5.3b}
\end{align}
where
\begin{equation} \label{3.5.4}
  \begin{aligned}
    U_0 &= (\alpha +\mathrm{i} \beta )^4 \left(1-12\Lambda \FF\left( \frac23 \Lambda \right)\right), \\
    U_1 &= 4 (\alpha +\mathrm{i} \beta )^3 (\alpha -\mathrm{i}\beta ) \left(1-12\Lambda \FF\left(\frac 23 \Lambda\right)\right), \\
    U_2 &= 6 \left(\alpha ^2+\beta ^2\right) \left((\alpha ^2+\beta ^2)\left(1+
   \frac {64}9\Lambda^2 \FF'\left(\frac 23\Lambda\right) -12 \Lambda \FF\left(\frac23 \Lambda\right)\right) -64 k \frac \Lambda3 \FF'\left(\frac 23\Lambda\right)\right), \\
    U_3 &=4 (\alpha +\mathrm{i} \beta ) (\alpha -\mathrm{i}\beta )^3 \left(1-12\Lambda \FF\left(\frac 23 \Lambda\right)\right), \\
    U_4 &=(\alpha -\mathrm{i} \beta )^4 \left(1-12\Lambda \FF\left(\frac23\Lambda\right)\right),
  \end{aligned}
\end{equation}

and
 \begin{equation} \label{3.5.5}
  \begin{aligned}
    V_0 &= (\alpha +\mathrm{i} \beta )^4 \left(1-12\Lambda \FF\left(\frac 23 \Lambda \right) \right), \\
    V_1 &= 4 (\alpha +\mathrm{i} \beta )^2 \Big((\alpha ^2+\beta ^2)\left(1+4\Lambda^2 \FF'\left(\frac 23 \Lambda\right) -6 \Lambda \FF\left(\frac 23 \Lambda\right)\right) \\
    &+6 k \left(\FF\left(\frac 23 \Lambda\right) -2\Lambda \FF'\left(\frac 23 \Lambda\right)\right) \Big), \\
    V_2 &=6 (\alpha ^2+\beta ^2) \Big((\alpha ^2+\beta^2)\left(1-\frac{16}9 \Lambda^2 \FF'\left(\frac 23 \Lambda\right) -4 \Lambda \FF\left(\frac 23\Lambda\right) \right) \\
    &+8 k \left(\FF\left(\frac 23 \Lambda\right) + \frac 23  \Lambda \FF'\left(\frac 23\Lambda\right) \right)\Big), \\
    V_3 &= 4 (\alpha -\mathrm{i} \beta )^2 \Big((\alpha ^2+\beta ^2)\left(1+4\Lambda^2 \FF'\left(\frac 23 \Lambda\right) -6 \Lambda \FF\left(\frac 23 \Lambda\right)\right) \\
    &+6 k \left(\FF\left(\frac 23 \Lambda\right) -2\Lambda \FF'\left(\frac 23 \Lambda\right)\right) \Big), \\
    V_4 &= (\alpha - \mathrm{i} \beta )^4 \left(1-12\Lambda \FF\left(\frac 23 \Lambda \right) \right).
  \end{aligned}
\end{equation}
We consider these equations as polynomials in $e^{2\mathrm{i} \sqrt{-\frac \Lambda3} t}$.
 It is clear that equations are satisfied for $\alpha^2 +\beta^2 = \frac {3k}\Lambda$. In the other case, $\alpha^2 +\beta^2 \neq \frac {3k}\Lambda$
 it remains to solve the following system of equations
\begin{equation}
  U_0 = U_1 = U_2 = U_3 = U_4 =0, \qquad V_0 = V_1 = V_2 = V_3 = V_4 =0.   \label{3.5.6}
\end{equation}

Equations of motion are satisfied in the following two nontrivial cases:
\begin{align}
(i): \quad  &\alpha^{2}+ \beta^{2} = \frac {3k}{\Lambda} , \label{3.5.7a} \\ % [1pt]
(ii): \quad &\mathcal{F}(\frac 23\Lambda)=\frac{1}{12 \Lambda} , \ \ \mathcal{F}\;'(\frac 23\Lambda) =0 , \ \  \alpha^{2} + \beta^{2}= -\frac {k}{\Lambda} .
\label{3.5.7b}
\end{align}

\noindent$(i)$:\quad In the first case, we have $R(t)= 4 \Lambda$.

\noindent$(ii)$:\quad In the second case, we have $R(t)=4 \Lambda + 8k a(t)^{-2}$. Taking $k= +1$, there exists $\varphi$ such that
\begin{align*}
\alpha= \frac{1}{\sqrt{-\Lambda}}\sin \varphi,\qquad\qquad
\beta= \frac{1}{\sqrt{-\Lambda}}\cos \varphi.
\end{align*}
Now, we can transform scale factor $a(t)= \alpha \cos \sqrt{-\frac{\Lambda}{3}}\;t + \beta \sin \sqrt{-\frac{\Lambda}{3}}\;t$ to
\begin{align}
a(t)=\frac{1}{\sqrt{-\Lambda}}\sin (\sqrt{-\frac{\Lambda}{3}} t- \varphi).  \label{3.5.8}
\end{align}

Effective energy density and pressure are:
\begin{align}
  \bar \rho = \frac{3k - \Lambda (\alpha^{2}+\beta^{2})}{8 \pi G \;a(t)^{2}}, \qquad\qquad
  \bar p =\frac{\Lambda(\alpha^{2}+\beta^{2})-3k}{24 \pi G \;a(t)^{2}}.  \label{3.5.9}
\end{align}
For $k \neq \frac{\Lambda}{3} (\alpha^{2}+\beta^{2})$ we have $\bar w =-\frac{1}{3}$.

\subsection{Cosmological Solutions of the Form $a_6(t) = \left(\alpha \cos 2\sqrt{-\frac{\Lambda}{3}}\ t +
\beta \sin 2\sqrt{-\frac{\Lambda}{3}}\ t\right)^{\frac 12}$}

In this case
\begin{equation}
  R=4\Lambda , \quad k= 0 . \label{3.6.1}
\end{equation}
From equations of motion follows
\begin{equation}
  \alpha^2 + \beta^2 = 0. \label{3.6.2}
\end{equation}
Hence, there are no nontrivial solutions of the form
 $$a_6(t) = \left(\alpha \cos 2\sqrt{-\frac{\Lambda}{3}}\ t +
\beta \sin 2\sqrt{-\frac{\Lambda}{3}}\ t\right)^{\frac 12} .$$

%%%%%%%%%%%%%%%%%%%%%%%%%%%%%%%%%%%%%%%%%%%%%%%%%%%%%%%%%%%%%%%%%%%%%%

\section{Discussion and Conclusions}

To have more complete presentation of the contents of this paper, some main considerations should be discussed. These considerations include
gained new cosmological solutions, used  eigenvalue method and nonlocal operator.

 \textit{On new  cosmological solutions.}
Section 3 is related to the finding of new cosmological solutions of nonlocal gravity model \eqref{eq2.1}.
In a class of possible scale factors of the form $ a(t) = (\alpha e^{\lambda t} + \beta e^{-\lambda t})^\gamma ,$ we have found  four new solutions
when $\gamma = 1 $ and no nontrivial solutions if $\gamma \neq 1$. The new solutions are:
\begin{align}
&a(t) = A\ e^{\pm \sqrt{\frac{\Lambda}{3}}\, t} , \quad R(t) = \frac{6k}{A^2} \ e^{\mp 2\sqrt{\frac{\Lambda}{3}}\, t} + 4\Lambda ,
\quad k= +1, -1 , \quad \Lambda > 0 . \label{eq4.1a} \\
&a(t) = \frac{1}{\sqrt{\Lambda}} \cosh\big(\sqrt{\frac{\Lambda}{3}} \, t\big) , \quad
  R(t) = 8 k\Lambda \frac{1}{\cosh^2\big(\sqrt{\frac{\Lambda}{3}} \, t\big)}  + 4\Lambda ,   \quad  k = -1 , \quad \Lambda > 0. \label{eq4.1b} \\
&a(t) = \frac{1}{\sqrt{\Lambda}} \sinh\big(\sqrt{\frac{\Lambda}{3}} \, t\big) , \quad
   R(t) = 8 k\Lambda \frac{1}{\sinh^2\big(\sqrt{\frac{\Lambda}{3}} \, t\big)}  + 4\Lambda ,   \quad  k = +1 , \quad \Lambda > 0 .  \label{eq4.1c} \\
&a(t)  = \frac{1}{\sqrt{- \Lambda}} \ \sin\big(\sqrt{\frac{-\Lambda}{3}}\ t\big) , \quad R(t) = - 8 k\Lambda
    \frac{1}{\sin^2\big(\sqrt{\frac{-\Lambda}{3}}\ t\big)}  + 4\Lambda, \quad k = +1 , \quad \Lambda < 0 . \label{eq4.1d}
\end{align}

Recall that in the de Sitter (anti-de Sitter) \eqref{eq2.5} case analogous solutions are:
\begin{align}
&a(t) = A\ e^{\pm \sqrt{\frac{\Lambda}{3}}\, t} , \quad R = 4\Lambda ,
\quad k= 0 , \quad \Lambda > 0 . \label{eq4.2a} \\
&a(t) = \sqrt{\frac{3}{\Lambda}} \cosh\big(\sqrt{\frac{\Lambda}{3}} \, t\big) , \quad
  R =  4\Lambda ,   \quad  k = +1 , \quad \Lambda > 0. \label{eq4.2b} \\
&a(t) = \sqrt{\frac{3}{\Lambda}} \sinh\big(\sqrt{\frac{\Lambda}{3}} \, t\big) , \quad
   R =  4\Lambda ,   \quad  k = -1 , \quad \Lambda > 0 . \label{eq4.2c} \\
&a(t)  = \sqrt{\frac{-3}{\Lambda}} \ \sin\big(\sqrt{\frac{-\Lambda}{3}}\ t\big) , \quad R =  4\Lambda, \quad k = -1 , \quad \Lambda < 0 . \label{eq4.2d}
\end{align}

Comparing \eqref{eq4.1a} -- \eqref{eq4.1d} with \eqref{eq4.2a} -- \eqref{eq4.2d} we can note that for the same cosmological constant $\Lambda$
there are analogous scale factors with the same time dependence, but with different curvature constant $k$.  This fact can be interpreted as change
of topology in  de Sitter (anti-de Sitter) space by inclusion of nonlocal term of the form $(R-4\Lambda)\mathcal{F}(\Box)(R-4\Lambda) ,$
see \eqref{eq2.1}. For example, exponential expansion \eqref{eq4.2a} in a flat de Sitter universe remains exponential \eqref{eq4.1a} by
nonlocal transition into  closed or open de Sitter space. We can also conclude that this kind of nonlocality changes  constant space-time curvature
($R = 4\Lambda$) to the time dependent one (R = R(t)).  It is worth noting that in  nonlocal square root gravity model \eqref{eq2.2} there is cosmological solution with the scale factor $a(t) = A  e^{\pm \sqrt{\frac{\Lambda}{6}}\ t}, \quad \Lambda > 0, \ k =  +1, -1,$ with scalar curvature
$R(t) = \frac{6k}{A^2}\ e^{\mp \sqrt{\frac{2}{3}\Lambda}\ t} + 2\Lambda$, see Sec. 3.3 in \cite{PLB}. This  case is similar to \eqref{eq4.1a} presented in this paper. We expect that analogous cases exist in some other examples of transition from local to nonlocal de Sitter model.

 \textit{On eigenvalue  method.}
In our approach, to solve equations of motion  in the case of a homogeneous and isotropic universe, essential role plays possibility to solve
the corresponding eigenvalue problem $\Box (R(t) - 4\Lambda) = q (R(t) - 4\Lambda) ,$ where $q = \zeta \Lambda .$ $\Lambda$ appears here
on the basis of dimensionality. Analogous solutions of \eqref{eq4.1a} -- \eqref{eq4.1d} and \eqref{eq4.2a} -- \eqref{eq4.2d} have the same Hubble
parameter $H(t) = \frac{\dot{a}}{a}$ and consequently the same d'Alembert-Beltrami operator $\Box = -\frac{\partial^2}{\partial t^2} -
3 H(t)\ \frac{\partial}{\partial t}$.
%Hence, analogous scalar curvatures ($R(t) = 4 \Lambda$ and $R(t) \neq 4 \Lambda$) can be regarded as
%as a result of different eigenfunctions $R(t) - 4\Lambda$ (with different eigenvalues) of the same operator $\Box$.

One can easily see that solution of $\Box (R - 4\Lambda) = q (R - 4\Lambda)$  implies solution of
the following eigenvalue problem:
\begin{align}
\Box^{-1} (R - 4\Lambda) = q^{-1} (R - 4\Lambda) , \quad q \neq 0 . \label{eq4.3}
\end{align} In other words, operators $\Box$ and $\Box^{-1}$ have the same eigenfunctions $R(t) - 4\Lambda$, but with different eigenvalues
$q$ and $1/q$, when $q \neq 0.$

 \textit{On nonlocal operator.}
Solvability of  eigenvalue problem \eqref{eq4.3} gives rise to introduce an extended version
of the nonlocal operator $\mathcal{F} (\Box) = \sum_{n = 1}^{+\infty}\ f_n \Box^n$ to the following one:
\begin{align}
\mathcal{F} (\Box) = \sum_{n = -\infty}^{+\infty}\ f_n\ \Box^n =  \sum_{n = 1}^{+\infty}\ f_n\ \Box^n  + f_0 + \sum_{n = 1}^{+\infty} f_{-n}\ \Box^{-n} ,
\label{eq4.4}
\end{align}
where $f_0 = 0$ in \eqref{eq2.1} nonlocal gravity model. Note that nonlocal operator \eqref{eq4.4} is symmetric under interchange $n \longleftrightarrow -n$.
This extended nonlocal operator satisfies eigenvalue problem $\mathcal{F} (\Box) (R - 4\Lambda) = \mathcal{F} (q) (R - 4\Lambda) ,$ where
\begin{align}
\mathcal{F} (q) = \sum_{n \neq 0}\ f_n\ q^n =  \sum_{n = 1}^{+\infty}\ f_n\ q^n  + \sum_{n = 1}^{+\infty} f_{-n}\ q^{-n} .
\label{eq4.4a}
\end{align}

In Section 3, we could replace $\mathcal{F} (\Box)$ by this one in \eqref{eq4.4} with $f_n = 0$, and the same new scale factors would
be obtained with the same constraints on this extended $\mathcal{F} (\Box)$. Note that now eigenvalues are: $q = \frac{2}{3} \Lambda, \ \ \Lambda \neq 0$ and
$q^{-1} = \frac{3}{2} \frac{1}{\Lambda}, \ \ \Lambda \neq 0$ for all four new solutions.

Note that finding of each new cosmological solution induces two restrictions on nonlocal operator $\mathcal{F} (\Box)$. At this stage an explicit form of
$\mathcal{F} (\Box)$ is not necessary.

\textit{On further investigations.}
Absence of the additional degrees of freedom, in particular ghosts, should be an important property of nonlocal gravity. A ghost-free
condition is investigated in paper \cite{dimitrijevic4} for  models of form \eqref{eq1.1}, which includes  our model \eqref{eq2.1}, see also
\cite{biswas6,buoninfante} and references therein.
To avoid a ghost, nonlocal
operator $\mathcal{F}$ must satisfy some conditions that depend on the background cosmological solution. This needs detailed investigation
of the second variation \cite{dimitrijevic9} of  action \eqref{eq1.1}  and is a subject for future consideration.

As it is shown in Section 2, nonlocal gravity model \eqref{eq2.1} can be derived from  nonlocal de Sitter gravity \eqref{eq2.2}. These two
models together contain cosmological solutions that mimic interference of dark energy with radiation and dark matter in the flat universe.
Both models also have a nonsingular bounce solution. Hence, at the cosmological scale, these nonlocal models imitate some effects that are a part of
cosmic history described by standard model of cosmology ($\Lambda$CDM model). This situation gives rise to continue with developments of this
nonlocal gravity approach and
explore influence on astrophysical effects at galactic scale  and the Solar system. It should be also investigated
possible inflation, cosmic microwave background (CMB) and cosmological perturbations \cite{mukhanov}.

{\it Conclusions.}
At the end, it is worth noting the main results presented in this paper.
\begin{itemize}
  \item Four new exact cosmological solutions are obtained.
  \item Effective energy density and effective pressure are computed for all new solutions.
  \item Change of space topology by nonlocal gravity is noted.
  \item A connection between nonlocal gravity models \eqref{eq2.1} and \eqref{eq2.2} is shown.
  \item Method of  finding eigenfunctions $R(t) - 4 \Lambda$ is further elaborated.
  \item  Nonlocal operator $\mathcal{F}(\Box)$ can be naturally extended by addition of $\Box^{-1}$ in a symmetric way.
\end{itemize}
\end{document}